# ANOMALOUS SKIN EFFECT STUDY OF SUPERCONDUCTING FILM

Binping Xiao[†], M. Blaskiewicz, T. Xin

*Brookhaven National Laboratory, Upton, New York 11973-5000, USA*

The field distribution inside the superconducting radiofrequency (SRF) film with different mean free path is studied using niobium (Nb) as an example. The surface resistance of clean Nb film with different substrate and different film thickness is calculated. We also show the study of a special structured multilayer superconducting film called Superconductor-Insulator-Superconductor (SIS) structure.

## I. Introduction

Recent advance in the surface treatment study pushes SRF cavities for accelerator made from bulk niobium (Nb) to its theoretical limit, both in the peak field and surface resistance [1-3]. And the long photon lifetime of SRF resonator makes it a superior choice for nonaccelerator low-field applications, i.e., serving as a key component in three-dimensional circuit QED architecture [4, 5]. Superconducting films are of great interest for both accelerator and nonaccelerator SRF applications. The history of using thin films in SRF application can be traced back to 1964 [6], at the very early stage of SRF technology. Candidate materials for this application include Pb, Nb, NbN, NbTiN, $Nb_3Sn$, $MgB_2$ etc, and numerous studies have been done to the SRF properties of these materials [6-28]. With the expansion of bulk and thin film SRF technology, however, there remains a need for an understanding of the RF field distribution along the depth of the superconductor. Previous studies assumed that the RF field decays exponentially inside the thin film [29, 30], this is true only for film with small mean free path $\ell$ [31-33]. In this study, we extend the study of the anomalous skin effect of normal conducting film to the case of superconducting [34]. Using Nb film as an example, RF field distribution inside film is presented; the surface impedances of films with different mean free paths, as well as different thicknesses, are calculated; Bulk Nb, or films with lossless dielectric substrate, and films with Cu substrate are studied; finally, we show the surface impedance of an interesting multilayer SIS structure [35]. This study is also applicable to cubit in which bulk or thin film superconductors are used [36]. In this study it is assumed that state-of-art technologies can be used to ensure nm range surface roughness both for substrate polishing and for film deposition [37, 38], so that the effect of surface roughness can be ignored.

## II. Anomalous skin effect of superconducting film

For bulk superconductor with diffuse reflection, its surface impedance can be calculated from [39, 40]:

$$Z = i\pi\omega\mu_0 \left\{ \int_0^\infty ln[1+K(p)/p^2]dp \right\}^{-1}$$

where

$$K(p) = \frac{-3}{4\pi\hbar v_F \lambda_L^2} \int_0^\infty \int_{-1}^1 e^{ipRu} e^{-\frac{R}{l}} (1-u^2)$$
$$\times I(\omega, R, T) du dR$$

and
$I(\omega, R, T)$
$$= -\pi i \int_{\Delta-\hbar\omega}^\infty [1-2f(E+\hbar\omega)][g\cos(\alpha\varepsilon_2) - i\sin(\alpha\varepsilon_2)]e^{i\alpha\varepsilon_1}dE$$
$$+\pi i \int_\Delta^\infty [1-2f(E)][g\cos(\alpha\varepsilon_1)+i\sin(\alpha\varepsilon_1)]e^{-i\alpha\varepsilon_2}dE$$

with

$$f(E) = \frac{1}{e^{\frac{E}{kT}}+1}$$

$$\varepsilon_1 = \sqrt{E^2 - \Delta^2}$$

$$\varepsilon_2 = \sqrt{(E+\hbar\omega)^2 - \Delta^2}$$

$$g = \frac{E(E+\hbar\omega) + \Delta^2}{\varepsilon_1 \varepsilon_2}$$

and

$$\alpha = \frac{R}{\hbar v_F}$$

$\lambda_L$ is the London penetration depth, $V_F$ is the Fermi velocity, $\Delta$ is the energy gap, $\hbar\omega$ is the photon energy, $T$ is the temperature, $R$ is the position, and $p$ is the momentum.

In order to calculate the RF field distribution within the film thickness $d$, a semi-infinite metal was considered, with its surface in the *xy*-plane and the positive z-axis directed towards the interior of the metal, and z=0 the interface between vacuum and metal. The electric field $E(z)e^{i\omega t}$ is in the x-direction. One needs to first do an inverse Fourier transform to $K(p)$ to get $k_a(z)$, and then apply it to a discrete equation, in which the film thickness $d$ (normalized to $\ell$) is divided into K equal segments $D = d/\ell/K$. Please note $K(p=0)$ is infinite and is replaced by $\frac{2d}{\pi}\int_0^{\pi/(2d)} K(p)dp$. This method was first crosschecked in normal conductor where both $K(p)$ and $k_a(z)$ have analytical expressions [34]. The electric field distribution $E_n$ can then be calculated from [34]:

$$E_n + \sum_{m=n}^{K}(n-m)[k^2\ell^2 E_m + i\omega\mu_0\ell^3 D^3 \sum_{j=0}^{K} E_j k_a((m-j)\ell)]$$
$$= E_d - \ell D(K-n)E_d'$$

[†]binping@bnl.gov

Here $k=\omega/c$, $E_d$ and $E_d'$ are $E$ and $E'$ at $z=d$. They are determined by the substrate with its surface impedance, noted as $Z_{sub}$, in the normal skin effect regime. The boundary condition is $E_d' = -i\omega\mu_0 E_d/Z_{sub}$. The above equation can be solved by matrix inversion to get $E_n$, with the results normalized to $E_d$ ($E_d = 1$), one exception is that for Perfect Electrical Conductor (PEC) boundary contion, $E_d = 0$ and the results are normalized to $E_d'$ ($E_d' = 1$). The surface impedance of the superconducting film can be calculated from the E field on the surface $z=0$ and its derivative over depth: $Z = -i\omega\mu_0 E_0/E_0'$.

For metal substrate with anomalous skin effect, one also needs to calculate $k_a(z)$ for the substrate and take into account the electrons coming out/going into the substrate. This case is not considered in this paper.

The material properties of Nb are listed in Table 1. We start with Nb without substrate (or with lossless dielectric substrate) in section III; and then discuss Nb film on Cu substrate in section IV with Cu in normal skin effect regime; and SIS structure with Nb film on lossless sapphire film on bulk Nb substrate is considered in section V. Convergence study has been performed for different film thicknesses, 6 steps per nm is used, for films thinner than 30nm, 200 steps are used.

Table 1. Material properties of Nb.

| $\lambda_L$ [nm] | 32 |
|---|---|
| $V_F$ [m/s] | $1.37\times10^6$ |
| $\Delta$ [meV] | 1.47 |

## III. Anomalous skin effect of Nb film without substrate

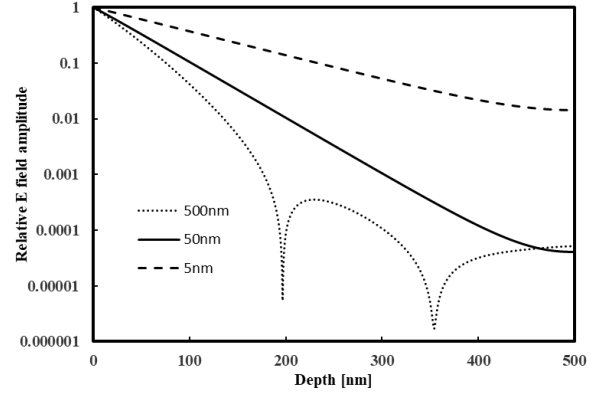

Figure 1. Amplitude of E field along the depth of 500nm thick Nb film at 1.5GHz and at 2K temperature without substrate, with 0 the interface between vacuum and film. Dotted line: with 500nm mean free path; Solid line: with 50nm mean free path; Dashed line: with 5nm mean free path.

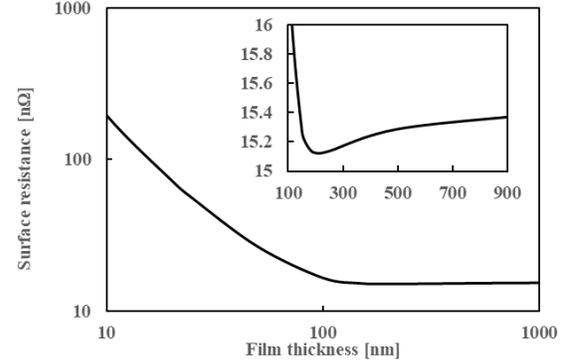

Figure 2. Surface resistance of Nb film versus film thickness for Nb with 500nm mean free path at 1.5GHz and at 2K temperature without substrate, with the inlet a zoom-in plot with thicknesses above 200nm.

Bulk Nb, which can be treated as Nb film without substrate, or Nb film with lossless dielectric material like $Al_2O_3$, is used to demonstrate the physical behavior of RF field in the film without considering the effect of the substrate. In this case the characteristic impedance of vacuum is used as the impedance of the substrate, $Z_{sub}$=376.73Ω. Nb films at 1.5GHz and at 2K temperature are studied. The E field amplitude along the depth of film is shown in Figure 1. From this plot one can notice that the E field amplitude decays monotonically along depth for Nb with 5nm or 50nm mean free path, and the decays are close to exponential. While for Nb with 500nm mean free path, it is not monotonic, and it deviates from exponential decay. Similar effect was also found in Cu film with anomalous skin effect, and the explanation is also similar, that there are energy exchanges between electromagnetic (EM) field and current carried by electrons, these energy exchanges are represented by the phase relationship between current and *E* field. For normal skin effect, they are in phase since $k_a(z)$ is a $\delta$ function, while for anomalous skin effect they are out of phase, causing this non-monotonic behavior [34]. The surface resistance vs different film thickness is plotted in Figure 2, from where one can see that a minimum surface resistance appears at 250nm thickness, and film thicker than 150nm should suffice. With really thin film (*d*<100nm), decreasing film thickness causes increasing surface resistance, which is because of the increasing in transmission to vacuum. With film thick enough, the surface resistance is close to the surface resistance of bulk Nb. The explanation of the minimum point is similar to the case of normal conducting film [34], that Nb film thickness smaller than the mean free path, but comparable with the effective penetration depth, which is between 40 and 500nm in this case, the diffussive reflection causes an enhancement in sheilding effect, thus a reduction in surface resistance while reducing the film thickness. Combining with the increasing in transmission to vacuum substrate which causes an increasing in surface resistance, a minimum appears. Please note in superconductor, the shielding effect is provided by Cooper pairs.

## IV. Anomalous skin effect of Nb film on Cu substrate

Using Nb film on Cu substrate to replace bulk Nb for SRF applications has many figure of merits including improving thermal conductivity, reducing acid usage in surface treatment, no need of electron beam welding, etc. It is cost effective, especially when high gradient is not required. At CERN, Nb film cavities are successfully adopted in LEP and LHC, with frequencies range from 350 to 500MHz at 4.3K temperature [8, 12, 41]. In this study, 400MHz frequency with 4.3K temperature is used. The surface impedance of Cu substrate is close to normal skin effect with surface resistance close to surface reactance, and $Z_{sub}$=(1.12+1.42*i*)mΩ. To clearly demonstrate the anomalous skin effect, Nb film mean free path $\ell$ =500nm is used, corresponding to RRR at around 50 [42], which requires advanced technologies, but is technically achievable [24, 43]. The surface resistance of Nb film with different thickness is shown in Figure 3, similar to the case of Nb film without substrate, a minimum surface resistance appears at 250nm thickness, and film thicker than 175nm should suffice. Note that both in Figure 2 and Figure 3, minimum surface resistance appears at similar film thickness, the reason is that the the penetration depth of Nb is not sensitive to the frequency, and it increases only a few percent from 2K to 4.3K.

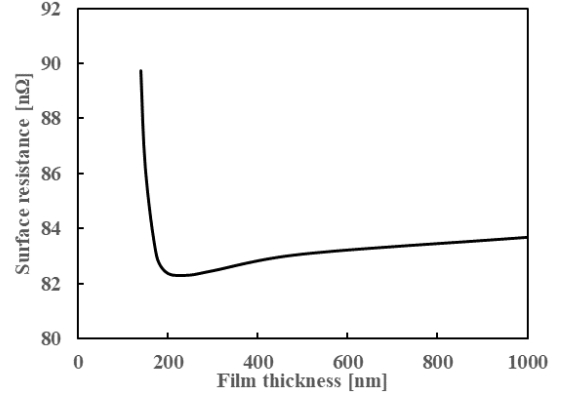

Figure 3. Surface resistance of Nb film with different thicknesses at 400MHz frequency and 4.3K temperature with 500nm mean free path.

## V. Anomalous skin effect of SIS multilayer film

The idea of using SIS multilayer films to enhance the peak magnetic field in SRF cavities was proposed by A. Gurevich [35]. It quickly became a hot topic in SRF community [44-53]. In this section, Nb-Sapphire-Nb structure is used as an example. At 1.5GHz and 2K temperature, the surface impedance of bulk Nb substrate is $Z_{bulk}$=(15.4+470000*i*)nΩ. Sapphire with dielectric constant 10.0, negligible loss tangent, and thickness $d_0$=15nm is used as the insulator layer, with its characteristic impedance $Z_I$=119.2Ω. Here tunneling effect is not considered since in tunneling junction <3nm thick insulator is normally used to preserve phase coherence across the superconducting layers [54]. This bi-layer (sapphire-bulk Nb) serves as a substrate for the top layer of Nb film with the substrate surface impedance to be:

$$Z_{sub} = Z_I \frac{Z_{bulk} + iZ_I Tan(d_0 \omega \mu_0 / Z_I)}{Z_I + iZ_{bulk} Tan(d_0 \omega \mu_0 / Z_I)}$$

based on transmission line theory [29]. In this study we use insulator with thickness in thens of nm. It is thin enough, resonance will not happen below 10GHz in the insulator layer while taking into account the reflection between Nb thin film and bulk Nb [29]. The surface impedance of this structure is not sensitive to the choice of the insulating material since with $Z_I$>1Ω, the value of $Z_{sub}$ is not sensitive to the value of $Z_I$. And with increasing insulator thickness, the real part of $Z_{sub}$ remains the same, while the imaginary part increasing.

The surface resistance of the SIS film with different top layer Nb thickness is plotted in Figure 4 top plot. For SIS film with 15nm sapphire, with the top Nb film layer thin enough, the surface resistances of this SIS film should be the same as the surface resistance of bulk Nb at 15.4nΩ, while the thickness gradually increases, the surface resistance gradually decreases slightly, to a local minimum at ~10nm thickness. With the thickness further increases, the surface resistance increases, with a peak of 16.3nΩ at 40nm thickness, close to the penetration depth. After that, the surface resistance decreases, to another local minimum at ~200nm. Then it slightly increases, with the surface resistance back to the bulk Nb value while the film is thick enough. The associated surface reactance is plotted in Figure 4 bottom plot. With film thick enough, its value is close to that of bulk Nb at 470μΩ. With film thin enough, its value is close to the value of the SI substrate, which is enhanced from the value of bulk Nb due to the transmission line effect of the sapphire, and the surface impedance gradually increases with decreasing film thickness. For comparison, the surface impedance of SIS film with a thicker sapphire at 30nm is also plotted in Figure 4, from where one can see that the surface reactance with film thin enough is bigger than that of film with 15nm sapphire, as we explained in the previous paragraph. Associating with this, a 17.8nΩ surface resistance peak at 30nm thickness appeared, higher than that of SIS film with 15nm sapphire.

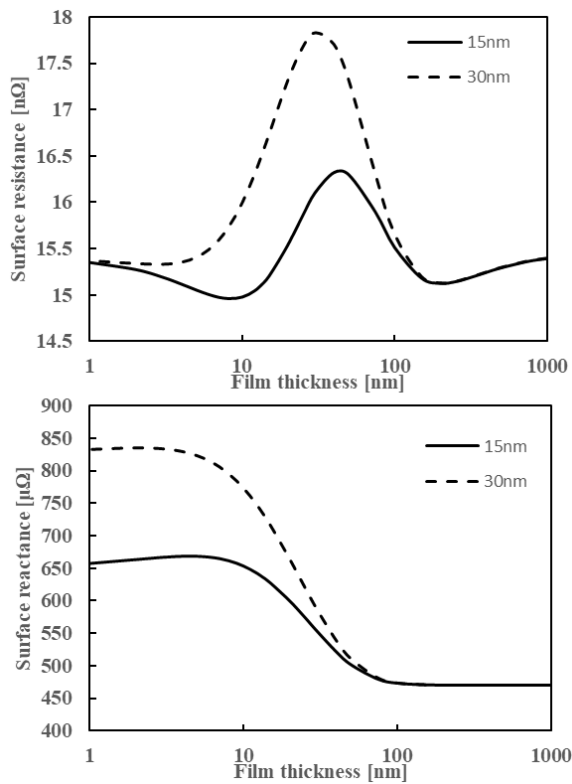

Figure 4. Surface impedance of SIS multilayer structure film, with thin film Nb on top with different film thicknesses, losses sapphire in the middle as the insulator layer, and bulk Nb on the bottom. Calculation is done at 1.5GHz frequency and 2K temperature with Nb 500nm mean free path. Surface resistance (top plot) and surface reactance (bottom plot) with different thickness for top layer Nb thin film is plotted. Solid line is for film with 15nm sapphire and dashed line is for film with 30nm sapphire.

## VI. Summary

Nb film without substrate was first studied in this paper, the amplitude of $E$ field along the depth of 500nm thick Nb film was found to be non-monotonic in clean films, and for dirty films the behave is close to normal skin effect with exponential decay. For clean Nb film without substrate, and clean Nb film with Cu substrate, minimum surface resistance is found at ~200nm film thickness, a number between penetration depth and mean free path, and it is not sensitive to resonant frequency. Clean Nb film with SIS structure is also studied, two local minima were found at ~10nm and ~200nm top layer film thickness, and maximum surface resistance was found at top layer film thickness around penetration depth, with its value increasing with increasing insulator thickness.

## ACKNOWLEDGEMENT

The work is supported by by Brookhaven Science Associates, LLC under contract No. DE-AC02-98CH10886 with the US DOE.